\documentclass[conference,a4paper]{IEEEtran}

\usepackage[pdftex]{graphicx}
\ifCLASSINFOpdf
\else
\fi

\usepackage{amsmath}

\ifCLASSOPTIONcompsoc
  \usepackage[caption=false,font=normalsize,labelfont=sf,textfont=sf]{subfig}
\else
  \usepackage[caption=false,font=footnotesize]{subfig}
\fi

\usepackage{url}

\newcommand{\argmin}{\mathop{\rm arg~min}\limits}

\usepackage{comment}

\begin{document}
\title{Home Location Estimation Using Weather Observation Data}

\author{
  \IEEEauthorblockN{Yuki Kondo\IEEEauthorrefmark{1} \quad Masatsugu Hangyo\IEEEauthorrefmark{2} \quad Mitsuo Yoshida\IEEEauthorrefmark{1} \quad Kyoji Umemura\IEEEauthorrefmark{1}}
  \IEEEauthorblockA{\IEEEauthorrefmark{1}Department of Computer Science and Engeneering,\\
  Toyohashi University of Technology,\\
  Aichi, Japan\\
  y153331@edu.tut.ac.jp, yoshida@cs.tut.ac.jp, umemura@tut.jp}
  \IEEEauthorblockA{\IEEEauthorrefmark{2}Weathernews Inc.,\\
  Chiba, Japan\\
  hangyo@wni.com}
}

\IEEEoverridecommandlockouts
\IEEEpubid{\makebox[\columnwidth]{978--1--5386--3001--3/17/\$31.00 \copyright 2017 IEEE \hfill} \hspace{\columnsep}\makebox[\columnwidth]{ }}

\maketitle

\begin{abstract}
We can extract useful information from social media data by adding the user's home location. 
However, since the user's home location is generally not publicly available, many researchers have been attempting to develop a more accurate home location estimation.
In this study, we propose a method to estimate a Twitter user's home location by using weather observation data from \textit{AMeDAS}.
In our method, we first estimate the weather of the area posted by an estimation target user by using the tweet,
Next, we check out the estimated weather against weather observation data, and narrow down the area posted by the user. 
Finally, the user's home location is estimated as which areas the user frequently posts from.
In our experiments, the results indicate that our method functions effectively and also demonstrate that accuracy improves under certain conditions.
\end{abstract}

\renewcommand\IEEEkeywordsname{Keywords}

\IEEEpeerreviewmaketitle

\section{Introduction}
Twitter\footnote{\url{https://twitter.com/} (accessed 2017--06--20)} is a social media that users can post short texts called \textit{tweets}.
There are many tweets posted on Twitter, and it can be regarded as a social sensor reflecting real world events~\cite{Atefeh15}.
Many researchers attempt to develop a more accurate user's home location estimation in order to use Twitter as a social sensor associated with location information~\cite{Zheng17}.
As methods of the user's home location estimation, for example, one using user's tweets posted by the user has been proposed.
Most of these methods (e.g.,~\cite{Cheng10,Kinsella11,Han12,Roller12}) estimate the user's home location by words that appear only in tweets of a few areas.
Such words are called \textit{local words}.
If most of the words included in user's tweets consist of local words in a certain area, it can be estimated that the user's home location is likely to be the area.
However, it is difficult to estimate the home location for users who have not posted such tweets containing local words.

In this paper, we focus on the tweet which contains weather information rather than one which contains local words.
For example, in the case of a tweet ``It is raining.'', the weather at the area the user posted can be regarded as rain at the posted date.
We propose a method to estimate the user's home location by using such tweets.
This method estimates the weather of the area posted by an estimation target user, then estimates the home location by using the estimated weather and the observed one.
In our experiments, the results indicate that our method functions effectively and also demonstrate that accuracy improves in rural areas.

\section{Related Work}
Many researchers have been attempting to develop methods in order to detect events and understand trends by considering Twitter as a social sensor~\cite{Atefeh15}.
Sakaki et al. detected an earthquake with high probability (96\% of earthquakes of JMA seismic intensity scale three or more are detected) merely by monitoring tweets~\cite{Sakaki10}.
Vieweg et al. reported situational features communicated in microblog posts during two concurrent emergency events that took place in North America during Spring 2009~\cite{Vieweg10}.
Aramaki et al. proposed a method to detect influenza epidemics by monitoring tweets and visualize the epidemic state\footnote{\url{http://mednlp.jp/influ_map/} (accessed 2017--06--20)}~\cite{Aramaki11}.
Herfort et al. presented a new approach that relies upon the geographical relations between twitter data and flood phenomena in order to enhance information extraction from social media~\cite{Herfort14}.
Asakura et al. collected posts concerning flood disasters from social media about weather and created a flood disaster corpus~\cite{Asakura16}.

We need the user's home location to use Twitter as social sensor.
However, user's home locations are generally not publicly available~\cite{Cheng10}.
Many researches attempt to estimate the user's home location using their information such as tweets~\cite{Zheng17}.
Most of these methods (e.g., \cite{Cheng10,Kinsella11,Han12,Roller12}) estimate the user's home location by local words included in the tweet.

Previous methods use words (local words) in which appearing areas are biased in order to estimate the user's home location.
However, there was a lack of geo-location information that can be used to extract such words.
In this paper, we propose a method to estimate the user's home location by using weather information, which can identify the actual area in advance.

\section{Proposed Method}
\subsection{Outline of Proposed Method}
In the proposed method, we first acquire relationships between the tweet and the weather by the following procedure.
First, tweets with location information are divided into words and are converted into feature vectors.
Next, each tweet is assigned a label as weather (raining or not) where the user posted by using weather observation data from AMeDAS.
Finally, we give feature vectors assigned labels to the support vector machine (SVM) and create a classifier for a tweet's weather estimation.

We collect tweets posted by an estimation target user and estimate labels using SVM against the tweets.
This label represents the weather where the estimation target user posted the tweet.
We regard the area where the estimated weather and the weather from AMeDAS data coincide as the user's home location.

\subsection{Classifier for Tweet's Weather Estimation}
First, we convert tweet $t$ to feature vector $\boldsymbol{x}_t$ by dividing the tweet into words as follows.
\begin{align*}
  \boldsymbol{x}_t&=(x_1, x_2, \ldots, x_{|W|}) \\
  x_i&=\begin{cases}
    1 \qquad w_i \in W_t \\
    0 \qquad (\text{otherwise})
  \end{cases}
\end{align*}
where $W$ is a set of words that occurred more than once in all the tweets,
$W_t$ is a set of words included in tweet $t$,
and $w_i$ is the $i$-th word when words included in $W$ are sorted in alphabetical order.
In this paper, we regard $W_t$ as all the words included in tweet $t$ under the limit of $W$.
Word selection for a better feature extraction is left for the future study.

Next, feature vector $\boldsymbol{x}_t$ is assigned a label as the weather where the user posted by using weather observation data from AMeDAS.
In label $l \in \{True, False\}$, $True$ indicates that it was raining at the posted area and $False$ indicates that it was not raining.

Finally, we give feature vectors assigned labels to the SVM as a state-of-the-art binary classifier.
In this paper, we used a linear kernel and set the cost parameter $C=0.025$ which sample value is shown in the scikit-learn document\footnote{\url{http://scikit-learn.org/stable/auto_examples/classification/plot_classifier_comparison.html} (accessed 2017--07--28)}.
We will plan to search for optimal parameters in the future study.

\subsection{Location Estimation by using Weather Observation}
In order to estimate a user's home location by using weather observation data,
we collect tweets posted by an estimation target user and estimate the labels (weather) of the tweets.
Thereafter, the user's home location is estimated based on the weather indicated by the estimated labels and the observed one.
In this paper, we hypothesized that a user would always posts tweets from the same area.

Let $T=\{t_1, t_2, \ldots, t_n\}$ be a set of tweets posted by an estimation target user and let $date(t_j)$ denote the posted date of the tweet $t_j \in T$.
Also, let $w\mathchar`- est(t_j)$ be an estimated weather of the tweet $t_j$.
In estimating weather, we estimate label $l \in \{True, False\}$ by using feature vector $\boldsymbol{x}_{t_j}$ converted from tweet $t_j$.
Then $w\mathchar`-est(t_j)\in \{True, False\}$.

Let $A=\{a_1, a_2, \ldots, a_m\}$ be a set of all areas, and we assume that a user's home location is one of $A$.
$m$ is the total number of areas, and each area corresponds one-to-one with the location of \textit{Meteorological Office} where AMeDAS is installed.
Let $w\mathchar`-obs(a,k)$ be the weather at area $a$ and date $k$, based on weather observation data from AMeDAS, then $w\mathchar`-obs(a,k)\in \{True, False\}$.
User's home location $a_{est}$ is estimated by following expression.
\begin{align*}
a_{est}&=\argmin_{a \in A} \textit{count}(T, a) \\
\textit{count}(T, a)&=\sum_{t \in T} \textit{diff}(w\mathchar`- est(t),w\mathchar`-obs(a, \textit{date}(t)))\\
\mathit{diff}(l_1, l_2)&=\begin{cases}
    0 \qquad l_1 = l_2 \\
    1 \qquad l_1 \neq l_2
  \end{cases}
\end{align*}

\section{Experiment Data}\label{dataset}
\subsection{Weather Observation Data}
AMeDAS (Automated Meteorological Data Acquisition System) is an automatic meteorological observing system operated by Japan Meteorological Agency (JMA).
AMeDAS is installed in about 1,300 locations throughout Japan and observes weather information such as precipitation, temperature and so on.
These weather observation data are published on the JMA's website.
However, the observation data of precipitation is recorded in units of 0.5 mm, and it is not possible to distinguish between precipitation less than 0.5 mm and no precipitation.
In other words, we cannot determine exactly whether it was raining or not.

In this paper, we use detailed rainfall observation data measured by a \textit{precipitation detector} and label whether it was raining or not.
A precipitation detector is a facility that detects slight precipitation of less than 0.5 mm, and is installed in 157 \textit{meteorological offices} throughout Japan.
As a result, it is possible to distinguish between precipitation less than 0.5 mm and no precipitation in the data observed at the meteorological office among the observation data from AMeDAS.

\subsection{Area and Tweet Data}\label{tweet-data}
Among the 157 meteorological offices listed on the JMA's website\footnote{\url{http://www.data.jma.go.jp/obd/stats/data/mdrr/chiten/sindex2.html}\\ \qquad(accessed 2017--04--20)}, we use 154 locations (areas) other than Showa Sta., Mt. Fuji and Mt. Aso.
Then $m=154$.
We assume that both locations where a tweet is posted and user's home locations are one of these areas.
We used the data on the latitude and longitude of the center of each area which is published on the JMA's website\footnote{\url{http://www.jma.go.jp/jma/kishou/know/amedas/kaisetsu.html}\\ \qquad(accessed 2017--04--20)}.

The following pre-processing applies to tweets with location information posted in 2016.
\begin{enumerate}
\item Remove tweets that have no location information consisting of latitude and longitude (``coordinates'') given by GPS.
\item Remove tweets posted by Bots judged based on client name (``source'').
\item Find the nearest AMeDAS within a 10 km radius based on the location information, and assign the area ID as the nearest AMeDAS to the tweet (remove the tweet if an area ID cannot be assigned).
\item Collate the weather observation data by the area ID and posted date, and assign a label as to whether it was raining or not to the tweet (remove the tweet if a label cannot be assigned).
\item Remove hashtags, URLs, screen names from the tweet.
\item Divide the tweet into words by using MeCab\footnote{\url{http://taku910.github.io/mecab/} (accessed 2017--06--20). ipadic-2.7.0.} if the tweet is in Japanese.
\item Remove tweets that contain none of the weather related words shown in Sec.~\ref{weather-words}.
\end{enumerate}
After the above pre-processing has been applied to the tweets, we assign the home location to users remaining over 10 tweets.
If 90\% or more of tweets are posted in the same area, we regard the area as the home location.
Tweets are removed for those users who cannot be assigned a home location.

In this paper, we pre-processed 171,614,139 tweets with location information posted in 2016, and eventually use 4,613 tweets posted by 153 users for our experiments.

\subsection{Collecting Weather-related Words}\label{weather-words}
Our method firstly estimates the weather of the area posted by an estimation target user, then estimates the home location by using the estimated weather and the observed one.
However, the weather of the tweet cannot be estimated if the tweet contains no weather information.
In this paper, we remove tweets that do not contain weather information before the experiment.

In order to remove tweets that do not contain weather information, we focused on words used to express weather in everyday conversation.
We use words such as ``sunny'' and ``good weather'' when expressing to ``It is sunny weather.'' in everyday conversation.
Therefore, if a tweet contains such words used to express a weather, it is considered that the tweet includes weather information at the posted area.
In this paper, we collect words used to express weather from tweets, and use only tweets containing collected words for our experiments.

In some cases, words used to express the weather appear unevenly for a day when the weather is either sunny or rainy.
Therefore, these words are considered to be biased to either $True$ or $False$ for label $l$ assigned to the tweet where the word appeared.

Let $W_t$ be a set of words included in tweet $t$.
Label $l$ is assigned to tweet $t$, although we assume that label $l$ is assigned to all words included in $W_t$.
We calculate the self-mutual information (PMI) of the following two events for each word and evaluate a bias of the word.
\begin{itemize}
\item The word appeared.
\item The label assigned to the word is $True$ (or $False$).
\end{itemize}
$\mathrm{PMI}(w,l)$ is the ease of attaching label $l$ to word $w$, this is defined as follows.
\begin{align*}
\mathrm{PMI}(w,l)&=\log{\frac{P(w,l)}{P(w)P(l)}} \\
P(w,l)&=\frac{C(w,l)}{N} \\
P(w)&=\frac{C(w)}{N} \\
P(l)&=\frac{C(l)}{N}
\end{align*}
where $C(w, l)$ is the number of times the label of word $w$ was $l$,
$C(w)$ is the total frequency of word $w$,
$C(l)$ is the total frequency of label $l$,
and $N$ is the sum of all word frequencies.

For the tweets set up to pre-process (5) in Sec.~\ref{tweet-data},
$\mathrm{PMI}(w,l)$ is calculated for the case of $l=\mathrm{True}$ and $l=\mathrm{False}$ for the word that appears more than 10 times,
and we collected 2,000 words with higher values.
However, there was a problem that low frequency words were included.
If the words are used by few users and the frequency is low, $P(w, True)$ and $P(w, False)$ values are greatly different and the PMI tends to be excessively high.
Therefore, we calculated the number of users using each word and removed words that ware used less than 10 times.
From the remaining words, we subjectively picked up weather-related words,
then we manually selected 50 words where the label tends to be $True$ and 18 words where the label tends to be $False$.

\section{Experiment Setting}
\subsection{Baseline Methods}
To verify the effectiveness of our method, we compare our method to baseline methods that use local words.
In order to compare in different viewpoints, we prepare two methods as the baseline methods.

Cheng et al.~\cite{Cheng10} propose a method for user's home location estimation by using the word occurrence probability of each area.
This method estimates the occurrence probability of words included in the user's tweets in each area, and consider the area where the sum of occurrence probabilities of words is maximum as the user's home location.
Let this method be the baseline~(A).

Our method consists of two steps, estimating weathers of user's tweets and estimating user's home location by using estimated weathers.
Therefore, we compare the estimation method that consists of two steps similar to our method.
For a set of words included in the tweet, the location where the tweet was posted can be estimated by the method similar to location estimation method of Cheng et al.~\cite{Cheng10}.
We construct the baseline~(B) as follows: estimate the locations where tweets were posted by an estimation target user, and estimate the user's home location by majority vote from the estimated tweet locations.

Although filtering and smoothing of words can be applied to both baseline methods, we do not apply them in our experiments.

\subsection{Evaluation Metrics}
In this experiment, we use $\mathrm{Precision}@k$ to evaluate the estimation performance.
This is the ratio of users where their actual home location is contained within the top $k$ estimated home locations.
We use the correct distance to evaluate whether an estimated home location is an actual home location.
This metric considers that the correct home location was estimated if the distance between an estimated home location and an actual home location is less than the threshold called correct distance.
The following expression defines $\mathrm{Precision}@k(U, d)$ where estimating home location for user set $U$ at correct distance $d$.
\begin{align*}
\begin{split}
\mathrm{Precision}@k&(U, d) = \\
&  |\{u|u \in U \wedge A_t(u, d) \neq \phi \}| / |U| \\
A_t&(u, d) = \\
& \{ a|a \in A_k(u) \wedge \mathit{dist}(a, a_t(u)) < d \}
\end{split}
\end{align*}
where $a_t(u)$ is the actual home location of user $u$,
$A_k(u)$ is a set consisting of the top $k$ estimated home locations for user $u$,
and $\mathit{dist}(a_1, a_2)$ is the distance between area $a_1$ and $a_2$.

$\mathrm{Precision}@k$ is also used for the evaluation of estimation performance by prefecture.
It differs in that user set $U$ is grouped by prefecture in which the actual user's home location belongs, and the precision is evaluated for each prefecture.
User set $U_p$ whose prefecture $p$ contains the user's actual home location is expressed as follows.
\begin{eqnarray*}
U_p = \{ u|u \in U \wedge \mathit{pref}(a_t(u))=p \}
\end{eqnarray*}
where $\mathit{pref}(a)$ is a prefecture that contains area $a$.
Let $P$ be a set of all prefectures in Japan,
for each $p \in P$, $\mathrm{Precision}@'(U_p, d)$ is calculated by the following formula.
\begin{align*}
\begin{split}
\mathrm{Precision}@k'(U_p, d)&= |\{ u|u \in U_p \wedge A_t'(u, d) \neq \phi \} | / |U| \\
A_t'(u, d)&= \{ a|a \in A_k(u) \wedge \\
\mathit{dist}(&a, a_t(u)) < d \wedge \mathit{pref}(a) = \mathit{pref}(a_t(u)) \}
\end{split}
\end{align*}

\section{Experiment Result}

\subsection{Evaluate Precision by Correct Distance}
\begin{figure}[!t]
\centering
\includegraphics[width=0.99\linewidth]{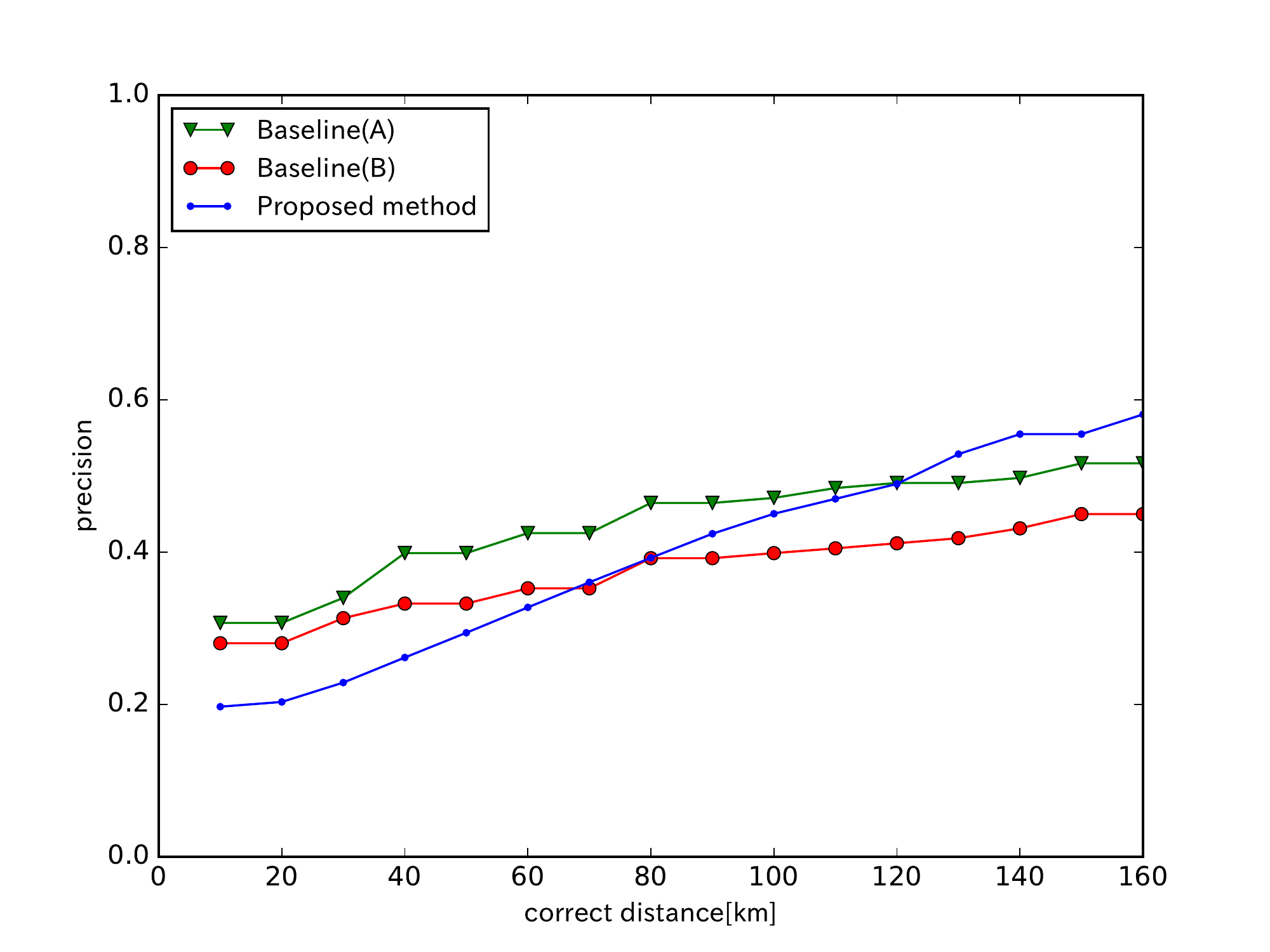}
\caption{Precision of Varied Correct Distance $(k=1)$.}
\label{fig:fig-result-k1}
\end{figure}

\begin{figure}[!t]
\centering
\includegraphics[width=0.99\linewidth]{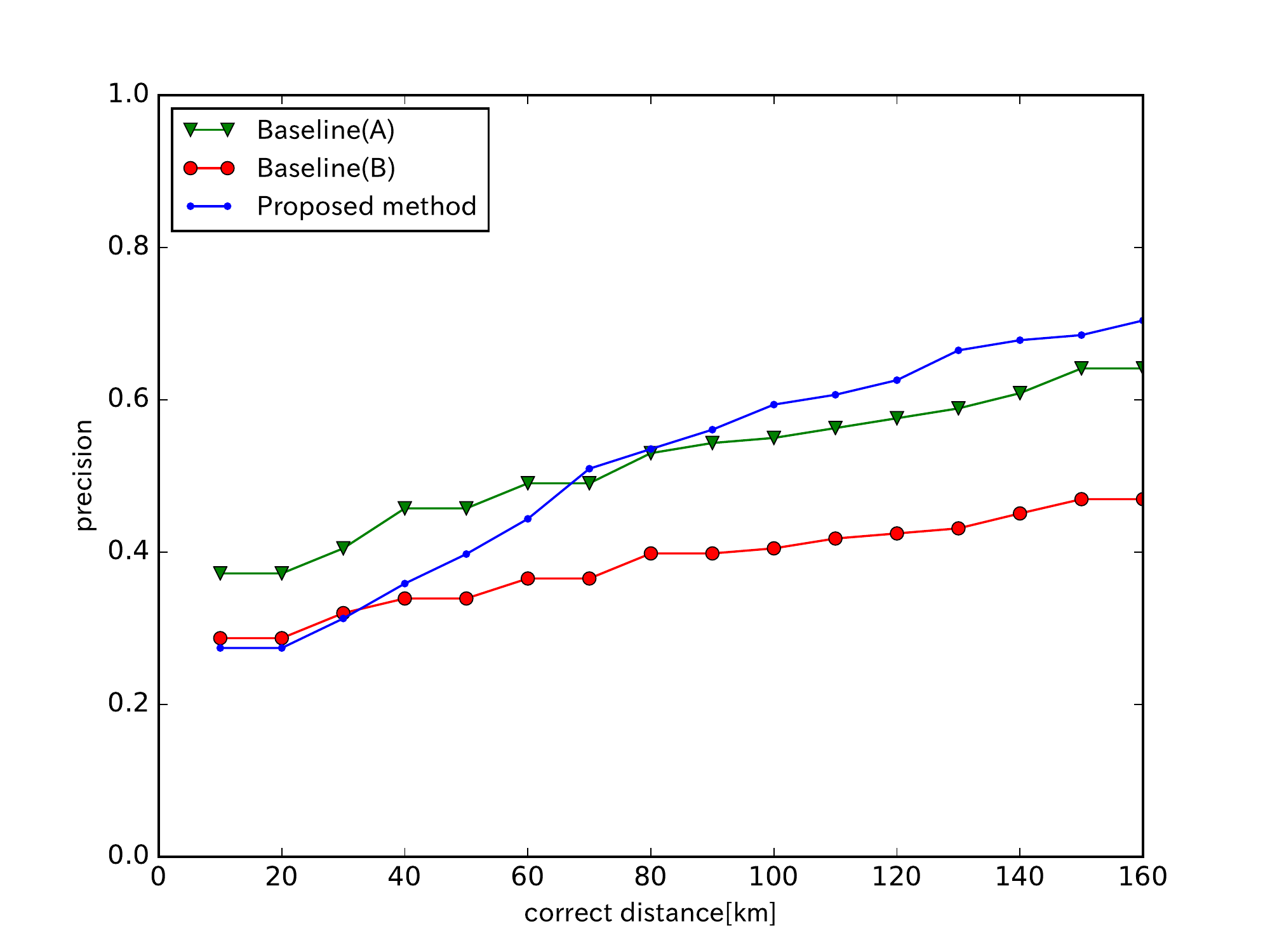}
\caption{Precision of Varied Correct Distance $(k=3)$.}
\label{fig:fig-result-k3}
\end{figure}

\begin{figure}[!t]
\centering
\includegraphics[width=0.99\linewidth]{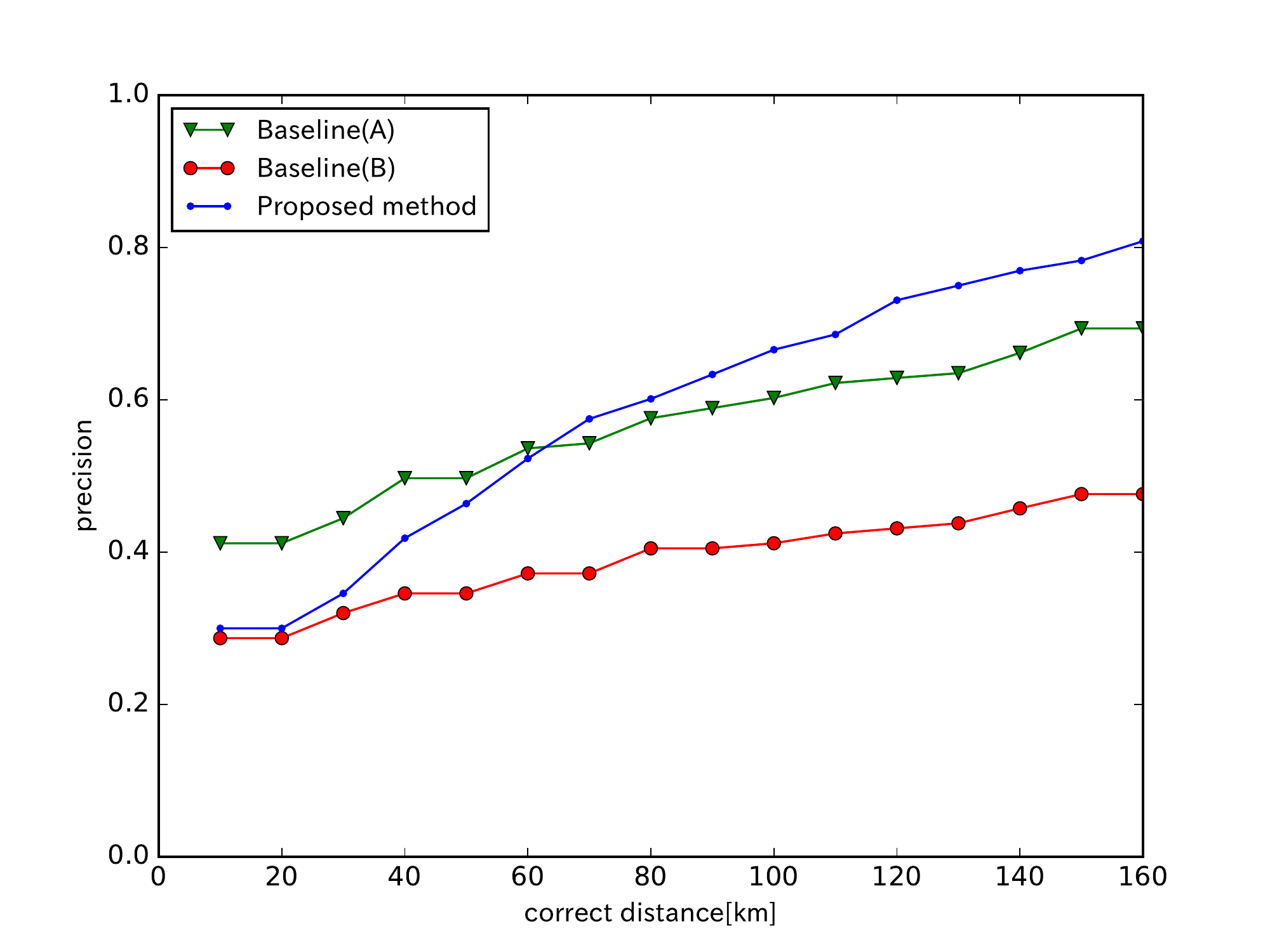}
\caption{Precision of Varied Correct Distance $(k=5)$.}
\label{fig:fig-result-k5}
\end{figure}

Fig.~\ref{fig:fig-result-k1}, \ref{fig:fig-result-k3} and \ref{fig:fig-result-k5} show $\mathrm{Precision}@k$ when $k =1,3,5$ and the correct distance $d$ is changed from 10 km to 160 km respectively.


\begin{figure*}[!t]
\centering
\includegraphics[width=0.99\linewidth]{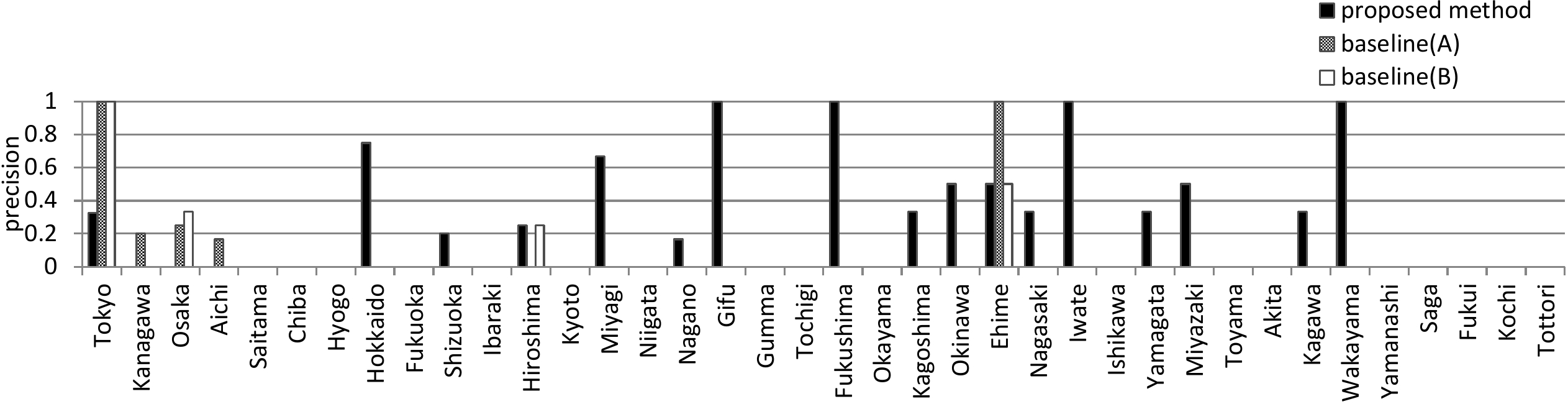}
\caption{\textbf{Precision for Each Prefecture:} Prefecture names are sorted in descending order of their population. The baseline methods can estimate correct locations for only a few prefectures that have a large population, whereas in our method it can estimate correct locations for prefectures with less population.}
\label{fig:fig-result-pref}
\end{figure*}

In all cases of $k = 1, 3, 5$, the baseline~(A) exceeds the baseline~(B).
Baseline~(B) is a method of estimating the posted location of tweets and estimating the home location by majority vote from the estimated tweet locations.
An error in posted location estimation of tweets is considered to have led to a reduction in $\mathrm{Precision}@k$.

Our method exceeds the precision of the baseline~(A) when the correct distance is larger than about 60 km to 80 km.
This is because our method can estimate correct locations for prefectures with less population as shown in Sec.~\ref{evaluate-prefecture}.
Also, the gradient of the graph is steeper than that of the baseline methods, and the improvement of the precision when the correct distance is larger is clearer.
In other words, it is indicated that our method estimated the area closer to the user's actual home location compared to the baseline methods even if it is incorrect.

\subsection{Evaluate Precision by Prefecture}\label{evaluate-prefecture}
Fig.~\ref{fig:fig-result-pref} shows $\mathrm{Precision}@k'$ evaluated by each prefecture with $k=1$ and correct distance $d=10$.
If there is no user in a prefecture, then the prefecture is not shown in Fig.~\ref{fig:fig-result-pref}.
In addition, prefecture names are sorted in descending order of their population in 2016\footnote{\url{https://www.e-stat.go.jp/SG1/estat/GL08020103.do?_toGL08020103_&listID=000001154737} (accessed 2017--05--15)}.
From this result, it is found that the baseline methods are correct only for prefectures with a large population.
In our method, accurate estimation is made in more prefectures than in the baseline methods.
This is clear from the macro average, and our method was 0.242, the baseline~(A) was 0.0689, and the baseline~(B) was 0.0458.
The macro average of our method is the largest.

Most location estimation methods use word frequencies for each area and estimate the occurrence probability of a word.
Some words have high occurrence probabilities in some areas, and such words are called local words.
Most location estimation methods use local words.
In Japan, however, it is known that about 40\% of tweets with location information were posted from the urban areas surrounding Tokyo, tweets are concentrated in urban areas.
Because the number of posted tweets depends on the area, words with equal occurrence probability in any area is estimated to have higher occurrence probability in areas with a large population.
Therefore, users who did not post tweets including local words are estimated to be that their home location is in an area with a large population.

According to experimental results, at the baseline methods, many users were estimated to be that their home location is in an area with a large population.
We used only tweets containing weather-related words for the experiments.
Therefore, this experimental setup is regarded as users that barely post tweets containing local words.
From the above discussion, our method is more effective than the baseline methods when users post many tweets that contain weather-related words not including local words.

\section{Conclusion}
In this paper, we have proposed a method to estimate the user's home location by using tweets without local words.
This method estimates the weather of the area posted by an estimation target user, then estimates the home location by using the estimated weather and the observed one.

We compared our method to the baseline methods that use local words.
As a result, under the condition of estimating three areas or more and the correct distance 70 km or more, the estimated precision of our method exceeded baseline methods.
In addition, we analyzed the estimated precision for each prefecture in order to find the factor that our method achieves more precision.
The baseline methods can estimate correct locations for only a few prefectures that have a large population, whereas in our method it can estimate correct locations for prefectures with less population.
From these experimental results, we found that in some cases baseline methods cannot estimate home location whereas our method can.

\bibliographystyle{IEEEtran}
\bibliography{IEEEabrv,references}

\end{document}